\documentclass{emulateapj}           
           
\shorttitle{Effect of photosphere} \shortauthors{Pe'er, M\'esz\'aros \& Rees}

\newcommand{\eV}{\rm{\, eV }}           
\newcommand{\keV}{\rm{\, keV }}           
\newcommand{\MeV}{\rm{\, MeV }}           
\newcommand{\GeV}{\rm{\, GeV }}

\newcommand{\llu}{\rm{\, erg \, s^{-1}}}           
\newcommand{\beq}{\begin{equation}}           
\newcommand{\eeq}{\end{equation}}           
\newcommand{\ba}{\begin{array}}           
\newcommand{\ea}{\end{array}}           
\renewcommand{\L}{L_{52}}          
\newcommand{\Gi}{\Gamma_{2}}          
          
\newcommand{\ee}{\epsilon_{e,-0.5}}          
\newcommand{\eB}{\epsilon_{B,-0.5}}          
\newcommand{\ed}{\epsilon_{d,-1}}          
          
\def \etal{{\it et al.~}}          
          
\begin{document}           
           
\title{The observable effects of a photospheric component on GRB's and XRF's prompt emission spectrum}           
           
\author{Asaf Pe'er\altaffilmark{1}\altaffilmark{2}, Peter M\'esz\'aros\altaffilmark{2} and Martin J. Rees\altaffilmark{3}}           
\altaffiltext{1}{Astronomical institute ``Anton Pannekoek'', Kruislaan 403, 1098SJ Amsterdam, the Netherlands; apeer@science.uva.nl}           
\altaffiltext{2}{Department of Astronomy and Astrophysics, Pennsylvania state university, University Park, PA 16802}               
\altaffiltext{3}{Institute of Astronomy, university of Cambridge, Madingley Rd., Cambridge CB3 0HA, UK}          
          
\begin{abstract}           
A thermal radiative component is likely to accompany the first stages          
of the prompt emission of Gamma-ray bursts (GRB's) and X-ray flashes          
(XRF's). We analyze the effect of such a component on the          
observable spectrum, assuming that the observable effects are due to a          
dissipation process occurring below or near the thermal photosphere.          
We consider both the internal shock model and a 'slow heating' model          
as possible dissipation mechanisms. For comparable energy densities in          
the thermal and the leptonic component, the dominant emission mechanism          
is Compton scattering. This leads to a nearly flat energy spectrum ($\nu F_\nu          
\propto \nu^0$) above the thermal peak at $\approx 10-100 \keV$           
and below $10-100 \MeV$, for a wide range of optical depths  $0.03          
\lesssim \tau_{\gamma e} \lesssim 100$, regardless of the details of          
the dissipation mechanism or the strength of the magnetic field.          
At lower energies steep slopes are expected, while above $100 \MeV$          
the spectrum depends on the details of the dissipation process.           
For higher values of the optical depth, a Wien peak is formed at $100          
\keV - 1 \MeV$, and no higher energy component exists.           
For any value of $\tau_{\gamma e}$, the number of pairs produced does          
not exceed the baryon related electrons by a factor larger than a few.          
We conclude that dissipation near the thermal photosphere can         
naturally explain both the steep slopes observed at low energies and         
a flat spectrum above $10 \keV$, thus providing an alternative         
scenario to the optically thin synchrotron - SSC model.          
\end{abstract}           
          

\section{Introduction}          
\label{sec:intro}          
The prompt emission from gamma-ray bursts (GRB's) is believed to arise         
from the prompt dissipation of a substantial fraction of the bulk kinetic        
energy of a relativistic outflow, originating from a central compact          
object \citep[for reviews, see][]{Mes02,Waxman03,Piran04}.           
The dissipated energy is converted to energetic electrons, which          
produce high energy photons by synchrotron radiation           
and inverse Compton (IC) scattering. This model was found consistent          
with a large number of GRB observations          
\citep{Band93,Tavani96,Preece98}.          
However, motivated by an increasing evidence for low energy spectral          
slopes steeper than the optically-thin synchrotron or synchrotron self          
Compton (SSC) model predictions \citep{Crider97, Preece98, Frontera00, GCG03},         
an additional thermal component was suggested which may contribute to          
the observed spectrum \citep{MR00,MRRZ,MR04}.           
Indeed, thermal radiation originating from the base of the          
relativistic flow where the densities are high enough to provide (at          
least approximate) thermal equilibrium is an inevitable ingredient.           
This radiation is advected outward as long as the flow (presumably,          
jetted) material remains opaque, and eventually emerges highly          
collimated once the flow becomes optically thin, at distance $r_{ph}$          
from the core, which defines the photospheric radius.          
The comoving energy density of this thermal radiation is comparable to          
the energy density of the relativistic outflow, in the first stages of         
the flow.          
             
The rapid variability observed in GRB's, if interpreted in the          
framework of the internal collisions model, implies that (at least          
some of)           
the internal collisions occur at radii smaller than the photosphere          
radius \citep{MR00, MR04}. A similar conclusion can be drawn if         
alternative energy dissipation mechanisms are considered, such as          
magnetic reconnection \citep{Thompson94, GS05}.          
If the energy dissipation occurs below or near the photospheric          
radius, the thermal photons serve as seed photons to IC scattering by          
the relativistic electrons. For high optical depth to scattering by          
the electrons (and the created pairs), the optically thin          
synchrotron-SSC model predictions are not valid, and a different          
analysis is needed.         
        
Saturated Comptonization of (synchrotron) soft photons         
was studied in the past in the context of X-rays emission from        
X-ray bursters \citep{ST80, PSS83, Zdziarski86, Zdziarski94}.      
It was first suggested as a source of        
GRB prompt emission by \citet{LKSC97}, and \citet{Liang97}.         
Comptonization by thermal population of electrons was considered by        
\citet{GC99} as an alternative source           
for GRB prompt emission. This model differs from the common        
``internal shock'' model in that it assumes that the dissipated kinetic          
energy is continuously distributed among all the shell electrons,       
rather than being deposited into a small fraction of electrons that          
pass through the shock wave, and it assumes that the energy is          
equally distributed among electrons rather than following a power law          
distribution. The implication of these models were further analyzed by          
\citet{PW04} \citep[see also][]{RR05}.           
          
In a previous paper \citep{PMR05a} we showed that clustering of          
the peak energy at $100 \keV - 1 \MeV$ as observed in many GRB's          
\citep{Brainerd98, Preece98a} can be naturally explained as being due        
to a Compton- inverse Compton scattering balance,          
if the kinetic energy dissipation process occurs below the Thompson          
photosphere. This result was found to be independent on the nature of the          
dissipation process.

In this paper, we analyze in greater depth the effect of a photospheric        
term on the prompt emission from GRB's and XRF's, under different          
assumptions about the unknown values of the free parameters of the          
emission models. In \S\ref{sec:photosphere} we show that the energy          
carried by the thermal photons is comparable to the dissipated energy        
for plausible values of the free parameters characterizing GRB's.       
This implies a large radiative efficiency, in contrast to conventional       
internal shock models.       
We analyze in \S\ref{sec:slow} the spectra obtained in the 'slow          
heating' scenario, and in \S\ref{sec:internal} the spectra obtained          
in the 'internal shock' scenario. We show that these two different         
scenarios lead to very similar spectra for a large range of parameters          
values. In particular, we show here that a flat energy per decade spectrum        
$\nu F_\nu \propto \nu^0$ arises naturally above a thermal peak at 30-100 keV.        
This provides a natural explanation for the ubiquitous $F_\nu \propto       
\nu^{-1}$ prompt emission spectra \citep{Band93, Preece00} in the sub-MeV to MeV     
range. At lower energies, the spectrum depends on the details of the scenario     
considered; it is generally steeper than the typical synchrotron spectrum,      
and can become flatter through superposition of different regions.     
In \S\ref{sec:numeric} we present numerical results, which         
confirm the validity of the analytical calculations in          
\S\ref{sec:slow}, \S\ref{sec:internal}, and provide further details of          
the spectra under various conditions.           
We discuss in \S\ref{sec:discussion} the implications of the possibility        
presented here of explaining the prompt emission by a scenario        
influenced by a photospheric component, different from the       
``standard'' optically thin synchrotron-SSC model, as well as  the        
applicability of this model to other compact objects, such as blazars.

\section{Energy dissipation and thermal component}          
\label{sec:photosphere}          
          
Following the analysis of \citet{MR04}, we assume that the          
photospheric radius $r_{ph}$, which is defined as the radius above          
which the comoving optical depth along the jet falls to unity, lies          
outside the saturation radius $r_s$, at which the bulk  Lorentz factor          
of the relativistic flow $\Gamma$ asymptotes to the dimensionless          
entropy $\eta =L_0/({\dot M} c^2)$,          
where $L_0$ and ${\dot M}$ are the total energy and mass outflow rates.           
The saturation radius is $r_s\sim r_0\eta$, where          
$r_0$ is the radius at the base of the flow (where the energy is          
released), which is comparable           
to the Schwarzschild radius $r_g$ of a central object of mass $M$ ,           
$r_0 \sim \alpha r_g =\alpha 2GM/c^2$  where $\alpha\geq 1$.          
Inside the saturation radius, the observer-frame photospheric luminosity           
$L_\gamma$ is approximately the total luminosity of the outflow $L_0$,           
since the increasing Doppler boost cancels the adiabatic decay of the           
comoving characteristic photon energy.           
At larger radius, $r>r_s$, the Lorentz factor no longer grows, and $L_\gamma           
(r)=L_0(r/r_s)^{-2/3} < L_0$, the greater part of the energy being in           
kinetic form, $L_k\sim L_0$.          
          
We assume the existence of a dissipation process, such as magnetic          
reconnection \citep{Thompson94, GS05}  or internal shock waves within          
the expanding wind \citep{RM94, WL95, SP97} that dissipates          
a fraction $\epsilon_d$ of the kinetic energy.           
As a concrete example for a dissipation process, we chose the internal          
shock model of GRB's, in which variation of the flow $\Delta \Gamma          
\sim \Gamma$ on a time scale $\Delta t \sim r_0/c$ results in the          
development of shocks at radius $r_i \approx 2 \Gamma r_s$.          
 At this radius, the photospheric luminosity is $L_\gamma          
(r_i)=L_0(2\Gamma)^{-2/3}$, and the comoving normalized temperature of          
 the thermal component is            
\beq          
\ba{ll}          
\theta & \equiv {k_B T' \over m_e c^2} = {k_B \over m_e c^2}\left( \frac{L_0}{4 \pi r_s^2 \Gamma^2 c a}          
  \right)^{1/4} ( 2 \Gamma)^{-2/3} \nonumber \\ & =1.2 \times 10^{-3} L_{52}^{1/4} \Gamma_2^{-5/3}          
  (\alpha m_1)^{-1/2},           
\label{eq:temp}          
\ea          
\eeq          
where $k_B$, $a$ are Bolzmann's and Stefan's constants, $L_0 = 10^{52}          
L_{52} {\rm erg\,s^{-1}}$, $\Gamma = 10^2           
\Gamma_2$, and the characteristic mass  $M\sim 10 m_1$ solar masses of          
the central object  (e.g. black hole) assumed, resulting in          
Schwarzschild radius  $r_g=2GM/c^2 \simeq 3\times 10^6 m_1~{\rm\,cm}$.

The comoving proton number density in the shocked plasma at $r_i$ is          
$n_p\approx L_0/(4\pi r_i^2 c \Gamma^2 m_p c^2)$, therefore the optical          
depth to Thompson scattering by the baryon related electrons at this          
radius is           
\beq          
\tau_{\gamma e} = \Gamma r_0 n_p \sigma_T = 100 \, L_{52}          
\Gamma_2^{-5} \alpha^{-1}.          
\label{eq:tau}          
\eeq          
We thus find that for the parameters values that characterizes GRB's,          
indeed $r_i < r_{ph}$.          
          
We assume that a fraction $\epsilon_d$ of the kinetic energy is          
dissipated by the dissipation process, resulting in an internal energy          
density $u_{int} = {L_0 \epsilon_d}/{4 \pi r_i^2 c \Gamma^2}$.            
Fractions $\epsilon_e$ and $\epsilon_B$ of this energy are carried by          
the electrons and the magnetic field, respectively, therefore the          
ratio of the thermal photon energy density $u_{ph}=aT'^4$           
to the electron energy density is           
\beq          
A \equiv {u_{ph} \over u_{el}} =  0.44 \Gi^{-2/3} \ed^{-1} \ee^{-1},          
\label{eq:epsilon_f}          
\eeq          
where $\epsilon_d = 10^{-1} \ed$, $\epsilon_e = 10^{-0.5} \ee$ and          
$u_{el} \equiv \epsilon_e u_{int}$.          
Since the thermal photons provide the main photon reservoir for          
electron scattering, the ratio of synchrotron and IC emitted power by          
the electrons is           
\beq          
S \equiv \frac{P_{syn}}{P_{IC}} = \frac{u_B}{u_{ph}} = 1.1          
\Gi^{2/3} \ed \eB,          
\label{eq:IC_syn}          
\eeq          
where $\epsilon_B = 10^{-0.5} \eB$, and $u_B \equiv \epsilon_B u_{int}$.

In order to determine the resulting spectrum, the particle energy          
distribution needs to be specified. In the analysis below, we consider           
two alternative scenarios: the ``slow heating'' scenario, and the more          
widely used internal shock scenario, in which power-law energy          
distribution of the injected electrons is assumed.

\section{Slow heating scenario}          
\label{sec:slow}          
          
In this scenario, as suggested by \citet{GC99}, the dissipated energy          
is continuously and evenly distributed among the shell's electrons,          
during the comoving dynamical time of the dissipation process,          
$t_{dyn} \equiv \Gamma \Delta t$.          
The electrons are assumed to have a Maxwellian distribution with          
normalized temperature $\theta_{el}$, and Comptonization is the main          
emission mechanism.          
          
The electron temperature is determined by energy balance between the          
energy injection rate, $dE_+/dt = u_{el}/ n_{el} t_{dyn}$, and energy          
loss rate. For optical depth $\tau_{\gamma e}$ not much larger than unity,          
the energy loss rate is given by $dE_-/dt = (4/3) c \sigma_T (\gamma_f          
\beta_f)^2 u_{ph} (1+S)$.          
 Here, $n_{el}$ is the number density of electrons in the plasma,          
$n_{el} = n_p$, pairs are neglected (see below), and $\gamma_f          
\beta_f$ is the electron momentum, which for a relativistic          
Maxwell-Bolzmann distribution is related to the temperature via          
$\theta_{el} = \gamma_f \beta_f^2 / (1 + \beta_f^2) $.          
At steady state, the electron momentum is therefore given by           
\beq          
\gamma_f \beta_f = \left({3 \over 4} {1 \over \tau_{\gamma e} A ( 1+          
S)}\right)^{1/2}.          
\label{eq:gbf_slow}          
\eeq          
Since $A$ and $(1+S)$ are of order unity for parameters characterizing          
GRB's, $\gamma_f \beta_f \simeq \tau_{\gamma e}^{-1/2}$.          
           
The resulting spectrum above the thermal peak can now be approximated          
in the following way.           
For $\gamma_f \beta_f \gtrsim 1$ ($\tau_{\gamma e} \lesssim 1$), the          
energy of a photon after $n_{sc}$ scattering is           
$\varepsilon_{n_{sc}} \approx (\gamma_f \beta_f)^{2n_{sc}}          
\varepsilon_0$, where $\varepsilon_0$ is the initial photons' energy.             
In the Thompson regime, the scattering rate is $dn_{ph}/dt \approx          
n_{ph,0} n_{el} c \sigma_T = n_{ph,0} \tau_{\gamma e}/ t_{dyn}$,          
where $n_{ph,0}$ is the number density of photons at the thermal peak.          
At the end of the dynamical time, the number density of photons that          
undergo $n_{sc}$ scattering is therefore $n_{ph,n_{sc}} \approx n_{ph,0}          
\tau_{\gamma e}^{n_{sc}}$. These photons have energies in the range          
$\varepsilon_{n_{sc}} .. \varepsilon_{n_{sc}} +          
d\varepsilon_{n_{sc}}$, therefore          
\beq     
\ba{lcl}
\log\left({n_{ph,n_{sc}} \over n_{ph,0}}\right) & =  &         
\log\left({\varepsilon_{n_{sc}} \over \varepsilon_0}\right)           
\times {\log \tau_{\gamma e} \over \log {(\gamma_f \beta_f)^2}} \nonumber \\ 
& = &             
\log\left({\varepsilon_{n_{sc}} \over \varepsilon_0}\right) \times (-1), 
\ea         
\eeq          
where $A$ not much different than 1 and $S$ not much larger than 1          
assumed in the last equality, resulting in $(\gamma_f \beta_f)^2          
\approx \tau_{\gamma e}^{-1}$.          
Since $n_{ph,n_{sc}} $ is the number density of photons in the range          
$\varepsilon_{n_{sc}} .. \varepsilon_{n_{sc}} +          
d\varepsilon_{n_{sc}}$, we conclude that $\varepsilon dn/d\varepsilon          
\propto \varepsilon^{-1}$, or $\nu F_\nu \propto \nu^0$ above the          
thermal peak and below $\varepsilon_{\max} = \gamma_f m_e c^2$          
(in the plasma frame).          
          
For $\tau_{\gamma e} \ll 1$, the electron momenta at steady state          
$\gamma_f \beta_f \gg 1$. An upper limit on the electron momenta          
$\gamma_f \beta_f \sim \gamma_f$  is $\gamma_f \leq \tilde{\gamma}          
\equiv u_{el} / n_{el} m_e c^2 = (m_p/m_e) \epsilon_d \epsilon_e          
\gamma_p = 60 \, \ed \ee \gamma_{p,0}$, where $(\gamma_p-1) m_p c^2$          
is the proton's thermal energy           
and $\gamma_p = 1 \gamma_{p,0}$ (cold protons) assumed.          
This value is achieved if the optical depth is $\tau_{\gamma e} \leq          
\hat\tau_{\gamma e} \approx 3           
\times 10^{-4} \, A_0^{-1} \ed^{-2} \ee^{-2} \gamma_{p,0}^{-2}$ where          
$A=1 A_0$, and equation \ref{eq:gbf_slow} was used. For lower value of          
$\tau_{\gamma e}$, the photons receive only a small fraction of the          
electron's energy.          
In this case only 1-2 scatterings are required to upscatter photons to          
$\epsilon_{\max} = \gamma_f m_e c^2$.  We therefore expect a spectrum          
with a wavy shape, composed of a low energy synchrotron peak,          
thermal peak, and few IC peaks (see the numerical results in          
\S\ref{sec:numeric} below).

For high value of the optical depth, $\tau_{\gamma e}$ larger than a        
few, the electrons momenta satisfy $\gamma_f \beta_f < 1$.           
If the optical depth is high enough, the photons receive nearly        
all of the electrons energy, and a Wien peak is formed at           
\beq          
\varepsilon_{WP} \approx 3 \theta m_e c^2 \times \left(1 + { 1 \over          
A} \right) \approx 6 \keV          
\label{eq:eps_wp}           
\eeq          
(in the plasma frame), where equations \ref{eq:temp}, \ref{eq:epsilon_f}          
were used.          
Electrons up-scatter photons up to energy $(\gamma_f \beta_f)^2          
m_e c^2$, therefore equation \ref{eq:eps_wp} sets a lower limit on          
$\gamma_f \beta_f$, $\gamma_f \beta_f \gtrsim 0.1$.           
Equation \ref{eq:gbf_slow} can now be used to find the value of          
$\tau_{\gamma e}$ above which the Wien peak is formed,           
$\tau_{\gamma e} > \tilde\tau_{\gamma e} \approx (3/4) (100/A(1+S))          
\approx 80 A_0^{-1}$.          
We thus conclude that for optical depth larger than $\sim 100$, the          
spectrum approaches a Wien spectrum, and equation \ref{eq:eps_wp}          
provides an asymptotic limit of the Wien peak energy.             
For $\tau_{\gamma e}$ of a few tens, the spectral slope below the          
Wien peak is softer than the Wien spectrum, $\nu F_\nu \propto          
\nu^\alpha$ with $\alpha \lesssim 3$. As a result the Wien peak is          
produced at energy somewhat higher than the asymptotic value given in          
equation \ref{eq:eps_wp}.

For values of $A$ not much smaller than 1, equation \ref{eq:gbf_slow}          
implies that a large number of pairs cannot be produced.          
If $\tau_{\gamma e}$ is larger than unity, $\gamma_f \beta_f < 1$,          
 the maximum energy of the up-scattered photons is smaller than          
$m_e c^2$, and no pairs are produced (for synchrotron emitted photons          
to be above $m_e c^2$, a magnetic field $B \gtrsim 10^{13}$~G is required).          
For $\tau_{\gamma e}$ smaller than 1, an upper limit on the number of          
the produced pairs can be obtained in the following way:           
For photon spectrum  $I_\nu \propto \nu^{-1}$,          
assuming the upscattering photons receive all the electrons energy,          
the number density of photons energetic than $m_ec^2$ is          
approximated by $n_\gamma(\varepsilon>m_e c^2) \approx u_{el}/m_e c^2          
\log(\gamma_f/3 \theta)$.            
The rate of pair production is $dn_\pm/dt \approx (3/16) c \sigma_T          
n_\gamma(\varepsilon>m_e c^2) n_\gamma(m_ec^2/\gamma_f \lesssim          
\varepsilon\lesssim m_ec^2)$,           
where $n_\gamma(m_ec^2/\gamma_f \lesssim \varepsilon\lesssim m_ec^2)$          
is the number density of           
photons that produce pairs with the energetic photons,          
$n_\gamma(m_ec^2/\gamma_f \lesssim \varepsilon\lesssim m_ec^2) \approx          
\gamma_f n_\gamma(\varepsilon>m_e c^2)$ for $I_\nu \propto \nu^{-1}$.          
The number density of the produced pairs at the end of the dynamical          
time is therefore          
$n_\pm \approx (3/16) c \sigma_T t_{dyn}\gamma_f u_{el}^2 /(m_e c^2          
\log(\gamma_f/3 \theta))^2$, or          
\beq          
{n_\pm \over n_{el}} = {3 \tilde{\gamma}^2 \gamma_f \tau_{\gamma e} \over 16          
\log\left({\gamma_f \over 3 \theta}\right)^2} \sim 6 \tau_{\gamma e}^{1/2},          
\label{eq:npm_ne}          
\eeq          
where characteristic value $\log(\gamma_f/3 \theta) \approx 10$          
assumed. For $\tau_{\gamma e}$ smaller than unity this result          
implies that pairs cannot outnumber the baryon related electrons by a          
factor larger than a few at most.           
For $\gamma_f \leq \tilde\gamma = 60$, a magnetic field larger than          
$\sim 10^{10}$~G           
is required for synchrotron emission at energies higher than $m_e c^2$,          
thus we conclude that pairs are not produced by synchrotron photons in          
this case of small $\tau_{\gamma e}$ as well.

\section{Internal collision scenario}          
\label{sec:internal}          
          
We further explore the effect of a photospheric component on the          
observed spectrum in the more conventional internal shocks scenario,          
in which the dissipated energy is converted to acceleration of          
electrons to high energies.           
We assume a fraction $\epsilon_{pl}\leq 1$ of the electron number          
density $n_{el}$ is accelerated to a power law energy distribution          
with power law index $p$ in the range $\gamma_{char}$ to $\gamma_{\max}$.          
A fraction $1-\epsilon_{pl}$ of the electrons are assumed to have a          
Maxwellian distribution with normalized temperature $\theta_{el} =          
\gamma_{char}/2$.              
Thermal photons are in the Klein-Nishina limit for Compton scattering          
by energetic electrons, therefore $\gamma_{max}$ is determined           
by equating the acceleration time and the synchrotron loss time,          
\beq          
\gamma_{max} = 2.3 \times 10^4 \L^{-1/4} \ed^{-1/4} \eB^{-1/4}          
\Gi^{3/2} \alpha_0^{1/2},          
\label{eq:gmax}          
\eeq          
where $\alpha = 1 \alpha_0$.          
The characteristic Lorentz factor of the accelerated electrons is           
\beq          
\ba{lcl}
\gamma_{char} & = & { \epsilon_e \epsilon_d \gamma_p \left(\frac{m_p}{m_e}\right)       
\over 3/2(1-\epsilon_{pl}) +          
\epsilon_{pl} \log\left(\frac{\gamma_{\max}}{\gamma_{char}}\right) } \nonumber \\      
& \simeq & 30 \ee \ed \gamma_{p,0},          
\label{eq:gmin}          
\ea
\eeq           
where $\epsilon_{pl} = 0.1$  was taken,          
$\log(\gamma_{\max}/\gamma_{char}) \approx 7$ and a power law index          
$p=2$ of the accelerated electrons above $\gamma_{char}$ is assumed.           
Electrons having Lorentz factor $\gamma$ lose their energy by Compton          
scattering and synchrotron emission on time scale           
$\tilde{t}_{loss} = {\gamma m_e c^2}/{(4/3) c \sigma_T \gamma^2          
u_{ph} (1+S)}$, and cool  down to $\gamma_f \simeq 1$ on time            
$t_{loss} = \tilde{t}_{loss}\times \gamma$. The ratio of the        
cooling time of electrons at $\gamma_{char}$ to the dynamical time is          
calculated using equations           
\ref{eq:tau},\ref{eq:epsilon_f}, \ref{eq:IC_syn} and \ref{eq:gmin},          
\beq          
{t_{loss}(\gamma_{char}) \over t_{dyn}} = {1 \over (4/3) A (1+S) \gamma_{char}          
\tau_{\gamma e}} ={ 0.03  \over \tau_{\gamma e}} \, \Gi^{2/3}.          
\label{tau_elec}          
\eeq           
We thus conclude that for Lorentz factor $\Gamma < 2\times 10^4$,          
$\tau_{\gamma e} > 1$ directly implies $t_{loss}(\gamma_{char}) <          
t_{dyn}$, and electrons lose all their energy during the          
dynamical time.

If the optical depth to scattering is low, $\tau_{\gamma e} < 0.03 \,          
\Gi^{2/3}$, electrons at $\gamma_{char}$ maintain their energy during          
the dynamical time. The resulting inverse Compton spectrum in the          
range $\theta m_e c^2 .. \gamma_{char}^2 \theta m_e c^2$ is $I_\nu          
\propto \nu$, or $\nu F_\nu \equiv \varepsilon^2          
dn_\gamma/d\varepsilon \propto \varepsilon^2$.           
At higher energies, the spectral index is determined by the power law          
index of the injected electrons. Above $\gamma_{char} m_e c^2$             
Klein Nishina effect significantly modifies the spectrum, and          
numerical treatment is required (see the numerical results in          
\S\ref{sec:numeric} below).

For an optical depth $0.03 \, \Gi^{2/3} \lesssim \tau_{\gamma e}          
\lesssim 1$, electrons lose their energy and accumulate at          
$\gamma_f \sim 1$ at the end of the dynamical time. The energy loss          
rate of electrons at Lorentz factor $\gamma< \gamma_{char}$            
is $p(\gamma) \propto \gamma^2 u_{ph} (1+S) \propto \gamma^2$,          
therefore the electron distribution below $\gamma_{char}$ is          
$n_{el}(\gamma) \propto \gamma^{-2}$, and the resulting IC spectrum in          
the range $\theta m_ec^2.. \gamma_{char} m_e c^2$ is           
$\nu F_\nu \equiv \varepsilon^2 dn_\gamma/d\varepsilon \propto          
\varepsilon^\alpha$ with $\alpha=1/2$.          
On top of the IC component there is the synchrotron component from          
the power-law accelerated electrons, with spectral index  $\alpha = 1- p/2$.          
The observed spectral index in this range is therefore expected to be          
$\alpha \lesssim 0.5$, the exact value depends on the details of the          
scenario, i.e. the values of $\epsilon_{pl}$, $S$ and $p$.           
In \S\ref{sec:numeric} we present our numerical results in this case.          
          
For higher optical depth, $\tau_{\gamma e} > 1$, electrons lose their          
energy by inverse Compton scattering and synchrotron emission, and          
accumulate at $\gamma_f \simeq 1$. At low energies, electrons are heated          
by direct Compton scattering energetic photons.           
We show in \S\ref{sec:appendix}, that a photon with energy $\varepsilon >          
f m_e c^2$ where $f \approx 3$ interacting with low energy electrons          
having $\gamma \simeq 1$, lose energy at an energy-independent          
rate, $d\varepsilon/dt \simeq c \sigma_T n_{el} (m_ec^2/2)$.          
The injection rate of energetic photons (via IC scattering and          
synchrotron emission) is approximated using energy considerations,          
$dn_{ph}(\varepsilon>f m_e c^2)/dt \approx u_{el}/f m_e c^2 t_{dyn}$.           
These photons lose energy by downscattering to energies below $f m_e          
c^2$ on a time scale $\varepsilon/(d\varepsilon/dt) = 2 f          
t_{dyn}/\tau_{\gamma e}$, therefore at steady state the number density          
of energetic photons is given by $n_{ph}(\varepsilon\gtrsim f m_e c^2)          
\approx 2 u_{el}/m_e c^2 \tau_{\gamma e}$.           
This result implies that the energy gain rate of electrons via          
direct-Compton scattering is equal to the energy gain rate in the slow          
heating mechanism, $dE_+/dt \simeq u_{el}/n_{el} t_{dyn}$          
\citep[see also][]{PMR05a}, therefore for $\tau_{\gamma e}$ not much          
larger than unity, equation \ref{eq:gbf_slow}          
characterizes the electron momenta at the end of the dynamical time          
in this scenario a well.           
We have previously shown \citep{PMR05a} that in this case of high        
optical depth, the electrons accumulate at $\gamma_f \beta_f \simeq  
0.1$ with a very weak dependence on the values of the unknown  
parameters, and that for optical depth higher than a few tens a Wien  
peak is formed at $1-10 \keV$. Equation \ref{eq:eps_wp} provides the  
asymptotic value of the Wien peak energy for very high optical depth  
in this scenario as well.            
For $\tau_{\gamma e}$ of a few tens, the thermal peak is upscattered to          
energy in the range $\theta m_e c^2 < \varepsilon_{peak} < \varepsilon_{WP}$          
(see the numerical results in \S\ref{sec:numeric}).          
          
In contrast to the slow heating scenario, a large number of pairs can          
in principle be created, following the injection of particles to          
high energies. However, the following argument, based on the          
analysis of \citet{PW04} suggests that the number of pairs do not          
outnumber the baryon-related electrons by a factor larger than a few,          
for any value of the optical depth $\tau_{\gamma e}$.          
From energy considerations, the pair injection rate is limited by          
$dn_\pm/dt \approx u_{el}/m_e c^2 t_{dyn}$. The pair annihilation rate          
is $\sim n_\pm^2 c \sigma_T$, therefore the ratio of pairs to          
baryon-related electrons in steady state is           
\beq          
{n_\pm \over n_{el}} \approx \left({\gamma_{char} \over \tau_{\gamma          
e}}\right)^{1/2}.          
\label{npm_nel_gen}          
\eeq          
We therefore do not expect pairs to outnumber the baryon-related          
electrons by a number larger than a few for any value of the optical          
depth. The creation of pairs lowers the number density of energetic          
photons, thus lowering the value of $\gamma_f \beta_f$ by a small        
factor  \citep[see][]{PMR05a}.

\section{Numerical Calculations}          
\label{sec:numeric}          
          
\subsection{The numerical model}          
\label{sec:model}          
          
In order to confirm the analytical calculations presented above and to          
derive the spectra resulting from different values of the free          
parameters, we calculated numerically the photon and particle energy          
distribution under the assumptions of \S\ref{sec:slow},          
\S\ref{sec:internal}.            
In our numerical calculations we used the time dependent numerical          
code presented in \citet{PW05}. This code solves self-consistently the           
kinetic equations for $e^\pm$ and photons describing cyclo-synchrotron          
emission, synchrotron self absorption, direct and inverse           
Compton scattering, pair production and annihilation, and the          
evolution of a high energy electro-magnetic cascade.          
These equations are being solved during the dynamical time of the          
dissipation process.

In this version of the code, we assume the          
existence of a thermal photospheric component, which at the          
dissipation radius $r_i$ is characterized by time independent luminosity           
$L_\gamma = L_0 (2 \Gamma)^{-2/3}$ and temperature given by equation          
\ref{eq:temp}.          
A fraction  $\epsilon_d = 0.1 \ed$ of the kinetic energy is          
dissipated by the dissipation process. A fraction $\epsilon_e$ of the          
dissipated energy is carried by the electrons, and a fraction          
$\epsilon_B$ is carried by the magnetic field.           
In the slow heating scenario, the dissipated energy is assumed to be          
continuously distributed among the electrons in the two shells (and the          
created pairs), which assume a Maxwellian distribution with          
normalized temperature $\theta_{el}(t)$. The electron temperature is          
determined self-consistently at each time step by balance of energy          
injection and energy loss.            
In the internal shock scenario, electrons are assumed to be injected          
by the shock waves at a constant rate. A fraction $\epsilon_{pl}$ of          
the electron population is injected into a power-law energy          
distribution with power-law index $p$ between $\gamma_{char}$ and          
$\gamma_{max}$ (see eqs. \ref{eq:gmax}, \ref{eq:gmin}). The remaining          
fraction $1-\epsilon_{pl}$ of the injected electrons assume a          
Maxwellian distribution with temperature $\theta_{el} = \gamma_{min}/2          
\approx 20 \ee \ed$.

\subsection{Slow heating scenario}          
\label{sec:slow_heat}          
          
Numerical results of the observed spectra at the end of the          
dissipation process in the slow heating scenario are presented in          
Figure \ref{fig:slow,tau} for different values of the optical depth          
$\tau_{\gamma e}$. In producing the plots in this graph, we chose           
representative values of the free parameters, $\alpha = 1$, $\Gamma =          
10^2$, $\epsilon_d = 10^{-1}$, $\epsilon_e = \epsilon_B =          
10^{-0.5}$. The differences in the plots are due to different intrinsic          
luminosity, $L_0 = 10^{48} - 10^{53}\llu$ which result in a wide range of          
optical depths, $10^{-2} \leq \tau_{\gamma e} \leq 10^3$ (see          
eq. \ref{eq:tau}). A similar graph is obtained  by assuming different          
value of $\Gamma$ and constant luminosity, where variation in the          
value of $\Gamma$ by a factor of 3, leads to variation in          
$\tau_{\gamma e}$ by a factor $10^{2.5}$.           
           
In the case of low optical depth $\tau<1$, one expects the thermal        
peak at the observed energy $3\theta \Gamma m_e c^2/(1+z) \approx 10-100        
\keV$, caused by the advected thermal photons.       
In the scenario of $\tau = 0.01$, the characteristic Lorentz factor          
of the electrons is $\gamma_f \approx 10$, and synchrotron radiation          
produces the peak observed at low energy, $\sim 300 $~eV.          
For higher optical depth the characteristic electron          
Lorentz factor is lower, $\gamma_f \gtrsim 1$ (see          
eq. \ref{eq:gbf_slow}), and the            
synchrotron emission occurs at lower energies, below the          
synchrotron self absorption frequency          
\citep[see the analysis of synchrotron self-absorption frequency as a          
  function of the free parameters values in][]{PW04}.           
          
The high value of the characteristic Lorentz factor of the electrons       
in the  scenario with $\tau = 0.01$ implies that two scatterings are        
required to upscatter thermal photons to $\sim 100 \MeV$, therefore Compton         
scattering of the synchrotron and thermal photons           
produces the wavy shape of the spectrum above 10 \keV.          
          
At moderate values of the optical depth, $\tau \sim 0.1 - 10$,          
multiple Compton scattering produces the flat spectrum observed above          
the thermal peak, in the range $10 \keV -  10 \MeV$, in agreement        
  with the predictions of \S\ref{sec:slow}.           
At even higher values of the optical depth,$\tau_{\gamma e} \gtrsim          
100$, a Wien peak is formed at          
sub \MeV energy. As explained in \citet{PW04}, the high optical          
depth implies that in this case photons are not observed prior to a period          
of adiabatic expansion, during which 50-70\%          
of the photons' energy is converted to kinetic energy (bulk          
motion) of the plasma. The observed Wien peak  is therefore expected          
at $\approx 300\keV$ \citep[see also][]{PMR05a}.

\subsection{Comparison of scenarios: low optical depth}          
\label{sec:low_tau}          
          
\begin{figure}          
\plotone{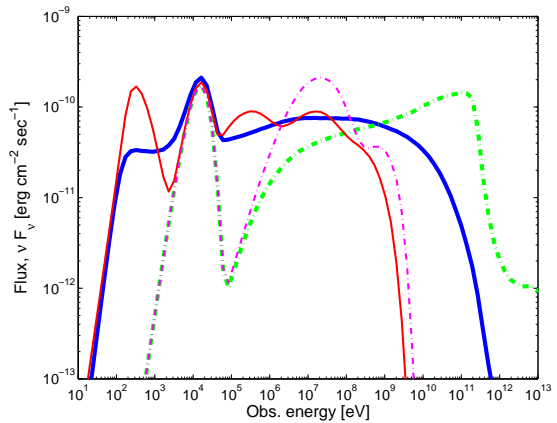}          
\caption{Time averaged spectra obtained for low value of          
 the optical depth, $\tau_{\gamma e} = 4\times 10^{-3}$. $\alpha = 1$,          
 $\L =10^{-2}$, $\Gamma = 300$, $\epsilon_d = 0.1$, $\epsilon_e =          
 10^{-0.5}$ assumed. Results for slow dissipation are shown in thin          
 lines and dissipation by shock waves with power law index $p=2$ and          
 $\epsilon_{pl} = 1$ are presented in thick lines.           
Solid lines are for high magnetic field, $\epsilon_B = 10^{-0.5}$ and          
 dash-dotted lines are for $\epsilon_B = 10^{-6}$. $z=1$  with the          
 same cosmological parameters as in figure \ref{fig:slow,tau}.            
 }          
\label{fig:tau_minimum}          
\end{figure}

Figure \ref{fig:tau_minimum} shows comparison of the spectra obtained          
for the two dissipation scenarios considered (slow heating and          
internal shocks) and two different values of the magnetic field, in the          
scenario of very low optical depth, $\tau_{\gamma e} = 4 \times          
10^{-3}$. For this value of the optical depth, only a small fraction          
of the electrons energy is radiated.         
If the magnetic field is weak (dash-dotted lines), the advected        
  thermal peak at  $10 \keV$ is prominent. In the slow heating        
  scenario (thin dash-dotted line) an IC peak at $\sim 15 \MeV$, created by        
  electrons having $\gamma_f \beta_f = 30$ is expected.            
The thin solid line, representing the slow          
heating scenario with equipartition magnetic field, shows four          
peaks: The synchrotron peak at $300 \eV$, the thermal peak at $10          
\keV$ and two peaks produced by IC scattering the low energy peaks, at          
$300 \keV$ and $10 \MeV$. The combined effect of these peaks is a wavy          
shaped, nearly flat spectrum in the range $100 \eV - 100 \MeV$.                
          
An internal shock scenario with weak magnetic field is shown by the          
thick dash-dotted line in Figure \ref{fig:tau_minimum}. Electrons          
injected with a power law index $p=2$ between $\gamma_{char}$ and          
$\gamma_{max}$ produce the high energy component by IC scattering the          
thermal photons.             
Electrons with Lorentz factor $\gamma < \bar\gamma = (4 \theta)^{-1} =          
3 \times 10^3$ cool by IC emission faster than the dynamical time,          
therefore the resulting IC spectrum is nearly flat, $\nu F_\nu \equiv          
\epsilon^2 dn_\gamma/d\epsilon \propto \epsilon^\alpha$  with $\alpha          
= 0$ below $\Gamma \bar\gamma^2 \theta m_e c^2/(1+z) \simeq 30 \GeV$.             
For energetic electrons with $\gamma > \bar\gamma$, the thermal          
photons are in the Klein-Nishina regime, and as a result these          
electrons cool slower than the dynamical time, and the spectral slope          
of the IC spectrum at higher energies is $\alpha = 1/2$.          
If magnetic field is added (thick, solid curve), electrons at all energies          
cool fast by synchrotron emission, and the resulting synchrotron          
spectrum is flat in the range $100 \eV - 10 \GeV$.

\subsection{Comparison of scenarios: intermediate and high optical depth}           
\label{sec:high_tau}          
          
\begin{figure}          
\plotone{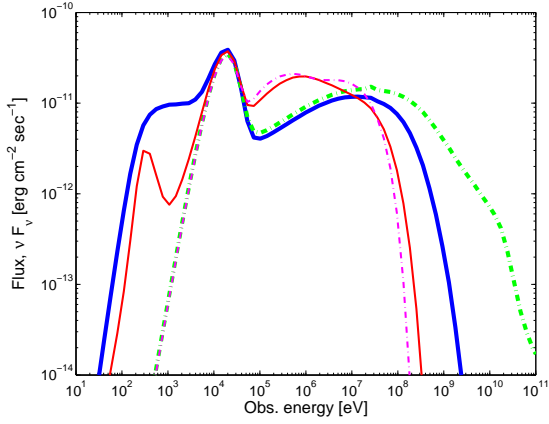}          
\caption{Time averaged spectra obtained for low value of          
 the optical depth, $\tau_{\gamma e} = 0.1$. $\alpha = 1$,          
 $\L =10^{-3}$, $\Gamma = 100$, $\epsilon_d = 0.1$, $\epsilon_e =          
 10^{-0.5}$ assumed.          
 Results for slow dissipation are shown in thin          
 lines and dissipation by shock waves with power law index $p=2$ are          
 presented in thick lines. Solid lines are for high magnetic field,          
 $\epsilon_B = 10^{-0.5}$ and dash-dotted lines are for $\epsilon_B =          
 10^{-6}$. $\epsilon_{pl} = 0$ (thick, solid), and $\epsilon_{pl} = 1$          
 (thick, dash-dotted). $z=1$  with the same cosmological parameters as          
 in figure \ref{fig:slow,tau}.             
}          
\label{fig:tau_0.1}          
\end{figure}

\begin{figure}          
\plotone{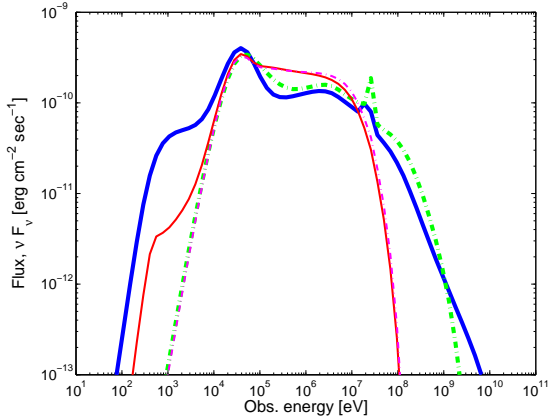}          
\caption{Time averaged spectra obtained for low value of          
 the optical depth, $\tau_{\gamma e} = 1$. Results are shown for           
 $\L =10^{-2}$ with all other parameters same as in          
 Fig. \ref{fig:tau_0.1}.           
 Results for slow dissipation are shown in thin          
 lines and dissipation by shock waves with power law index $p=2$ are          
 presented in thick lines. Solid lines are for high magnetic field,          
 $\epsilon_B = 10^{-0.5}$ and dash-dotted lines are for $\epsilon_B =          
 10^{-6}$. $\epsilon_{pl} = 1$ (thick, solid), and $\epsilon_{pl} =          
 0$ (thick, dash-dotted).            
}          
\label{fig:tau_1}          
\end{figure}

For low to moderate values of $\tau_{\gamma e}$, equation        
\ref{eq:gbf_slow} provides a good estimate of the electron momenta.         
The ratio between the energy of the synchrotron peak        
$\varepsilon_{syn}$ and the energy of the thermal peak          
$\varepsilon_{th} = 3 \theta m_e c^2$ is given by           
\begin{eqnarray}          
{ \varepsilon_{syn} \over \varepsilon_{th}} & = & {3 \over 2} {\hbar q B          
  \over m_e c} {(\gamma_f \beta_f)^2 \over \varepsilon_{th}} \nonumber \\
& = &            
\left({\pi a \over 8}\right)^{1/2} {2^{1/3} \over k_B^2} {\hbar q \over m_e c}          
{\Gamma^{1/3} (\epsilon_B \epsilon_d)^{1/2} \varepsilon_{th} \over A          
  (1+S) \tau_{\gamma e}} \nonumber \\          
& = & {4.5 \times 10^{-4} \over \tau_{\gamma_e}} \quad \Gi^{1/3} \,          
\ed^{1/2} \, \eB^{1/2} \, A_0^{-1} \, \varepsilon^{ob.}_{th,4},           
\end{eqnarray}          
where $\varepsilon^{ob}_{th} = 10^4 \varepsilon^{ob}_{th,4} \eV$ is the          
observed energy of the thermal peak.           
We therefore conclude that for $\tau_{\gamma e} \gtrsim 0.1$ the          
peak energy of the synchrotron component is much lower than the        
  peak energy of the thermal component, and is below the self absorption          
frequency for $\tau_{\gamma e} \gtrsim 1$.         
This conclusion is confirmed by the numerical results of the slow        
  heating scenario (thin line) with strong (solid) magnetic field, presented        
  in  Figures \ref{fig:tau_0.1} and \ref{fig:tau_1} for         
$\tau_{\gamma e} = 0.1,1$ respectively. A low energy synchrotron        
  component at $\sim 200 \eV$ is prominent for $\tau_{\gamma e} = 0.1$ (Figure        
\ref{fig:tau_0.1}), and is much weaker for  $\tau_{\gamma e} = 1$ (Figure        
\ref{fig:tau_1}).

\begin{figure}          
\plotone{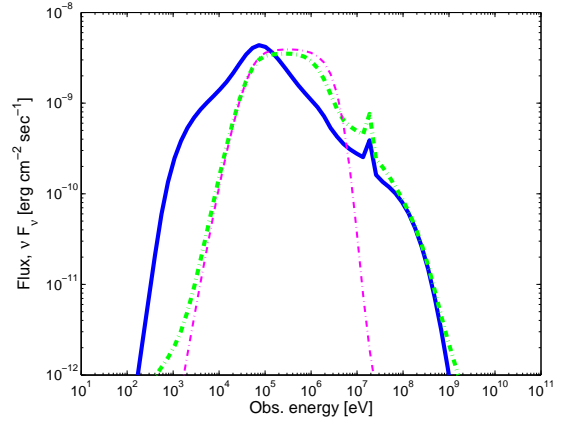}          
\caption{Time averaged spectra obtained for intermediate value of          
 the optical depth, $\tau_{\gamma e} = 10$. Results are shown for           
 $\L =10^{-1}$ with all other parameters same as in          
 Fig. \ref{fig:tau_0.1}.           
 Results for slow dissipation are shown in thin          
 lines and dissipation by shock waves with power law index $p=2$ are          
 presented in thick lines. Thick, solid line: high magnetic field,          
 $\epsilon_B = 10^{-0.5}$ and $\epsilon_{pl} = 0$. Thick, dash-dotted          
 line: $\epsilon_B = 10^{-4}$, $\epsilon_{pl} = 1$. Thin, dash-dotted:          
 slow heating, $\epsilon_B = 10^{-6}$.          
}          
\label{fig:tau_10}          
\end{figure}

Figures \ref{fig:tau_0.1}-\ref{fig:tau_10} provide a comparison of the       
  spectra obtained for the two dissipation mechanisms, equipartition        
  vs. low magnetic field, and two different values of $\epsilon_{pl}$.         
The internal shock scenario is shown with the thick lines in          
Figure \ref{fig:tau_0.1}. The solid line denote the scenario of equipartition          
magnetic field and $\epsilon_{pl} = 0$, i.e., the electrons are          
injected at $\gamma_{char}$ and than cool to $\gamma_f$.         
Synchrotron emission dominates the spectrum below the thermal        
  peak, at $0.1 -10 \keV$, and IC scattering dominates at higher energies.       
The electron distribution below $\gamma_{char}$ is $n_{el}(\gamma) \propto          
\gamma^{-2}$, therefore the emitted radiation has spectral slope          
$\alpha = 1/2$. IC scattering by electrons accumulated at $\gamma_f$          
results in somewhat flatter ($\alpha \lesssim 1/2$) spectral slope.          
The thick dash-dotted curve shows a scenario with weak        
magnetic field ($\epsilon_B = 10^{-6}$) and $\epsilon_{pl} = 1$.         
In this scenario, energetic electrons upscatter photons to        
high energies; however, pair production suppresses emission above        
$\GeV$.        
The number of pairs created in this scenario is found numerically to        
be $n_{\pm}/n_{el} \simeq 4$, in agreement with the analytical        
predictions of \S\ref{sec:internal}.          
          
The numerical results obtained for $\tau_{\gamma e} = 1,10 $ are          
presented in Figures \ref{fig:tau_1} and \ref{fig:tau_10}.         
These results further show the similarity between the spectra obtained        
by the two different dissipation mechanisms, as well as the weak        
dependence of the spectral shape on the value of the magnetic field       
or on $\epsilon_{pl}$.       
Multiple scattering creates the flat energy per decade       
spectrum above the thermal peak at $3 \theta \Gamma m_e c^2 / (1+z)      
\simeq 30 - 100 \keV$, for a typical $\Gamma=100-300$. This energy is      
very similar, albeit somewhat lower than the observed spectral break energy,      
$\sim 300 \keV$ \citep{Band93, Preece98a, Preece00}.     
The flat spectrum extends up to $\Gamma (\gamma_f \beta_f) m_e      
c^2/(1+z) \approx 10 \MeV$.      
      
In the internal shock scenario, the energetic photons produce          
pairs, resulting in a cutoff above $\sim 100 \MeV$. The number density          
of pairs in the scenario $\tau_{\gamma e} = 1$ is found numerically to        
be $n_{\pm}/n_{el} \simeq 4$ ($\epsilon_{pl} = 0$), $n_{\pm}/n_{el}        
\simeq 5$ ($\epsilon_{pl} = 1$).         
Similar, but somewhat lower values ($n_{\pm}/n_{el} = 1.5, 3.5$,          
respectively) are obtained for $\tau_{\gamma e} = 10$. We thus find       
that, indeed, pairs do not outnumber the baryon related electrons          
by a large factor, due to pair annihilation.          
The pair annihilation phenomenon causes the small peak observed at        
$\Gamma m_e c^2/(1+z) = 25 \MeV$ \citep[see][for further discussion]{PW04}.         
In summary, for intermediate values of the optical depth         
$\tau_{\gamma e} \approx 0.1 -10$, multiple IC scattering  produces an       
approximately flat energy per decade spectrum in the range        
$\approx \keV$ - sub \GeV for the two dissipation mechanism considered        
regardless of  the values of the magnetic field or of $\epsilon_{pl}$.            
          
\begin{figure}          
\plotone{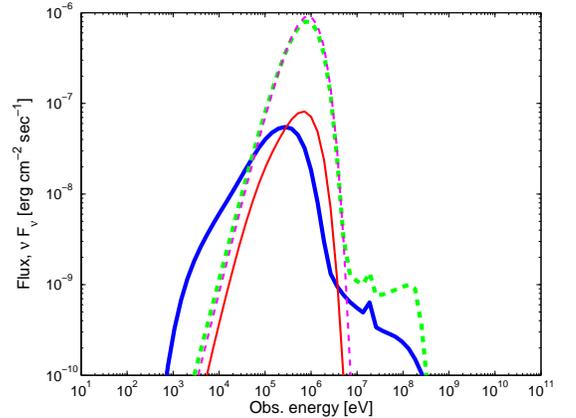}          
\caption{Time averaged spectra obtained for high value of the optical           
  depth, $\tau_{\gamma e} = 100,1000$. Results are shown for $\L          
  =1$ (solid), $\L = 10$ (dash-dash) with all other parameters          
  same as in  Fig. \ref{fig:tau_0.1}.            
Results for slow dissipation are shown in thin lines and dissipation          
  by shock waves with power law index $p=2$ are presented in thick lines.           
Thick, solid line: high magnetic field,$\epsilon_B = 10^{-0.5}$ and          
  $\epsilon_{pl} = 1$. Thick, dash-dash line: $\epsilon_B = 10^{-6}$,          
  $\epsilon_{pl} = 0.1$. Thin, solid: slow heating, $\epsilon_B =          
  10^{-0.5}$. Thin, dash-dash: slow heating, $\epsilon_B = 10^{-6}$.           
}          
\label{fig:tau_100}          
\end{figure}

Figure \ref{fig:tau_100} shows the asymptotic case of high optical          
depth, $\tau_{\gamma e} = 100, 1000$. For optical depth larger than          
$\sim 100$, multiple Compton scattering leads to the formation of          
a Wien peak at sub-\MeV energies, regardless of           
the details of the acceleration mechanism, or the values of any of          
the free parameters \citep[see also][]{PMR05a}.             
As shown in \S\ref{sec:internal}, for high values of the optical        
depth the number density of pairs is strongly suppressed by pair          
annihilation. This calculation is confirmed by the numerical results,          
 $n_{\pm}/n_{el} \simeq 1$.

\begin{figure}          
\plotone{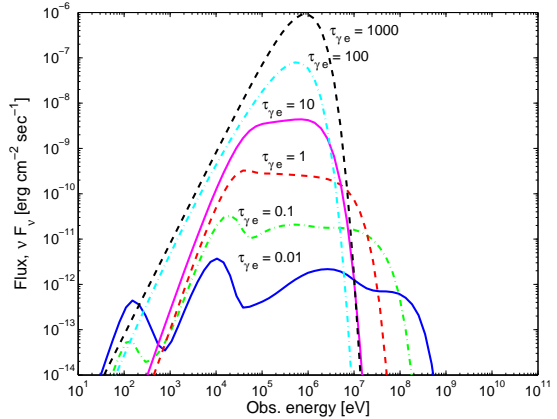}          
\caption{Time averaged spectra obtained for different values of          
  optical depth, $\tau_{\gamma e}$ in the slow heating          
  scenario. Results are for $\alpha=1$, $\Gamma = 100$,$\epsilon_d =          
  0.1$, $\epsilon_e = 10^{-0.5}$ and GRB luminosity $\L = 10^{-4},          
  10^{-3}, 10^{-2}, 10^{-1}, 10^0, 10^1, 10^2$, with corresponding          
  optical depth $\tau_{\gamma e} = 10^{-2} - 10^3$ (see          
  eq. \ref{eq:tau}). $\epsilon_B = 10^{-0.5}$ for $\L = 10^{-4} -          
  10^0$, and $\epsilon_B = 10^{-6}$ for $\L = 10^1, 10^2$. A redshift          
  $z=1$ in a flat universe with $\Omega_m = 0.3$, $\Omega_\Lambda          
  =0.7$, $H_0 = 70$ is assumed. The spectrum is not corrected for the          
  energy loss due to adiabatic expansion following the dissipation,          
  which lowers the energy for the cases of $\L = 10^1, 10^2$          
  ($\tau_{\gamma e} = 10^2, 10^3$) by a factor of 2-3.            
}          
\label{fig:slow,tau}          
\end{figure}          
          
\section{Summary and Discussion}          
\label{sec:discussion}          
          
In this work, we have considered the effect of a photospheric          
component on the observed spectrum after kinetic energy dissipation          
that occurs below or near the thermal photosphere of the outflows in       
GRB's or XRF's, with particular emphasis on the resulting spectral shape.         
Two dissipation mechanisms were considered: the ``slow heating''          
mechanism, and the internal shock scenario. We showed that the          
resulting spectra are largely independent on the details of          
the dissipation mechanism or of the values of any of the free parameters.          
A strong dependence was found only on the value of the          
optical depth, $\tau_{\gamma e}$. For $\tau_{\gamma e}$ smaller than a          
few tens, an approximately flat energy per decade spectrum is        
obtained above a break energy at few tens - hundred keV and up     
to sub GeV, while steep slopes are obtained below the break energy (see      
Figures \ref{fig:tau_1}, \ref{fig:tau_10}).     
For higher values of the optical depth,  a Wien peak is formed at        
sub\MeV energies, with steep slope at lower energies, and sharp cutoff      
above this energy.       
The production of pairs does not change this result, since pair      
annihilation limits the ratio of the number density of pairs to      
baryon-related electrons to a factor not larger than a few.          
          
Our theoretical and numerical results emphasize the important role of         
inverse Compton scattering in the formation of the spectra in scenarios         
involving an important thermal component.          
As a consequence, the observed spectral slopes in the two model considered          
are significantly different than the synchrotron model prediction. The          
(nearly) thermal component observed for low to intermediate      
values of the optical depth, and the Wien peak formed for high value      
of the optical depth can account for the increasing evidence for steep      
spectral slopes observed below $\sim 100 \keV$ in the early stages of GRB's          
\citep{Preece02,GCG03,Ryde04}.            
As shown in \citet{PMR05a} and as confirmed here (figures  
\ref{fig:slow,tau}, \ref{fig:tau_100}), for a scattering optical depth
larger than $\sim 100$, dissipation at a sub-photospheric radius can
naturally lead to a clustering of the peak energy at $\sim 100 \keV -
1 \MeV$, as observed in many bursts \citep{Brainerd98,Preece98a, Preece00}. 
 
For low values of the optical depth, $\tau_{\gamma e}\sim 0.1 - 1$,    
the advected thermal component produces an observed peak at a few   
tens of keV. This peak energy is consistent with the peak energy    
observed in X-ray flashes at $\sim 25 \keV$ \citep{Heise01}    
and X-ray-rich GRB's (XRR's), which show peak energy clustering at    
$\sim 50 \keV$ \citep{Sakamoto05}. Thermal peak energies in this range of    
few tens of keV are obtained for a wide range of values of the free    
model parameters, $10^{-4}\lesssim L_{52}\lesssim 1$, $1\lesssim    
\Gamma_2\lesssim 3$ and $1\lesssim \alpha \lesssim 10^3$ which result   
in an optical depth $\sim 1$ (see eqs.\ref{eq:temp},\ref{eq:tau}).     
It is difficult to obtain in our model a thermal peak energy at    
energies below a few keV combined with an optical depth larger than    
$\sim 10^{-3}$. Therefore, for a peak below a few keV, electrons    
maintain their energy (see \S\ref{sec:internal}). We conclude that if    
XRF's peak energies are due to advected thermal component, no XRF's    
with peak energy below a few keV are expected.     
   
The inclusion of an advected thermal component in the prompt emission
calculations can help resolving the question of the accelerated
electron distribution, which is of high theoretical importance. 
As pointed out by \citet{BB04}, fitting the data of many GRB's by
synchrotron and Compton emission only, require
the acceleration of the majority of electrons to a power law
distribution, leaving far too few thermal particles than to be
expected as a seed distribution for the power law population.
Here, we overcome this problem by assuming that the thermal component
seen in many bursts \citep[e.g.,][]{Ryde04}, and presumably exist in
all of them at early times, is advected from the core.
As a result, a variety of spectra can be obtained under different
assumptions on the values of the free parameters, without the
requirement that most of the 
electrons are accelerated to a power law distribution 
(in the presented figures of the power law acceleration model, we
allow the parameter $\epsilon_{pl}$ to vary between 0 and 1).

The spectra presented in this paper (\S\ref{sec:numeric}) describe the          
emission resulting from a single dissipation phase (e.g., single          
collision between two shells) for a particular choice of model          
parameters. Observed spectra are expected to be combination of spectra          
produced by many such dissipation processes, which are          
characterized by different parameters, e.g., different locations of          
collisions within the expanding wind. A detailed comparison with          
observations therefore requires a detailed model describing the          
distribution of single dissipation parameters within the fireball wind             
 \citep[see, e.g.,][]{DM98,PSM99,GSW01}. The construction and          
 investigation of such detailed models would involve additional assumptions      
 and parameters, which is beyond the scope of this manuscript.    
      
Nonetheless, the similarities found between the resulting spectra,      
produced under reasonable assumptions about the unknown values of the      
free parameters, suggests that the key results would remain valid for      
a spectrum that is obtained from a series of dissipation processes.      
These results include an observed break energy at few tens of keV to sub MeV,      
accompanied by slopes steeper than the optically thin synchrotron      
spectrum at lower energies, which are obtained for a single dissipation     
event for any value of the optical depth, and flat energy per decade      
spectra obtained up to sub-GeV energy for optical depths of  
  $10^{-2}$ - few tens. For a superposition of dissipation events at  
various radii, flatter slopes can be obtained.     
  
A superposition of dissipation events is also required in our model in    
order to explain both peak clustering at $\sim 300 \keV$ and flat    
energy per decade spectra at higher energies, as observed in many  
GRB's. A high optical depth $\tau_{\gamma e} \gtrsim 100$ is required
in order to obtain a Wien peak at sub-MeV energies, while lower values
of $\tau_{\gamma e}$ are required in order to obtain a flat energy per
decade spectra at higher energies. As detailed in \citep{PW04,
  PMR05a}, for the case of high optical depths the photons are
expected to lose $\sim 30\%$ of their energy to the bulk motion of the
plasma during an adiabatic expansion phase before escaping, thus the
observed Wien peak energy is expected at $\sim 300 \keV$.  
We conclude that given such a superposition, a clustering of the peak   
energy, steep slopes below the peak and flat energy per decade spectra 
above the break are a natural consequence of a model containing a    
significant thermal emission component.      
These results show good agreement with a large number of observations      
of GRB prompt emission spectra \citep{Band93, Preece00, Ryde04}.

The fluxes predicted by the model are within the detection          
capability of the Swift\footnote{http://www.swift.psu.edu} satellite          
in the \keV range, and the          
GLAST\footnote{http://www-glast.stanford.edu} satellite in the           
sub-\GeV range. Observations of a cutoff at high energies will therefore         
provide information about the optical depth during the dissipation          
phase, hence constraining the value of $\Gamma$, one of the least          
constrained free parameters of the model.           
          
The results may be applicable to a variety of compact objects,          
such as GRB's, XRF's and blazars. At least 60 flat spectrum          
radio quasars having flat ($\alpha \approx 0$) spectral index in the          
EGRET range were reported \citep[e.g.,][]{Mukher97}. Similar spectra were          
obtained for several quasars using the BeppoSAX satellite          
\citep{Tavecchio00}, while clustering of blazars peak energies at 1-5          
\MeV was reported by \citet{McNaron95}. We thus conclude that similar           
mechanisms to the ones presented here may explain some of the          
observable effects in radio quasars as well.           
          
\acknowledgments          
AP wishes to thank R.A.M.J. Wijers for useful discussions.

\appendix          
\section{Energy loss rate of high energy photons due to Compton          
scattering with mildly relativistic electrons}          
\label{sec:appendix}          
          
The rate of scattering by a single electron having Lorentz factor          
$\gamma$ passing through space filled with a unit density,          
isotropically distributed, mono-energetic photons with          
energy $\alpha_1 m_ec^2$ was first derived by \citet{Jones68},          
and summarized in \citet{PW05},          
\beq          
\frac{d^2N(\gamma,\alpha_1)}{dt d\alpha} =           
\frac{ \pi r_0^2 c \alpha}{2 \gamma^4 \beta \alpha_1^2}          
\left[F(\zeta_+) - F(\zeta_-) \right].          
\label{eq:compton_rate_general}          
\eeq          
Here, $\alpha$ is the energy of the outgoing photon in units of          
$m_e c^2$, $r_0$ is the classical electron radius, $\beta =          
(1-1/\gamma^2)^{1/2}$, $\zeta_\pm$          
are the upper and lower integration limits          
and $F(\zeta)$ is given by the sum of 12 functions, $F(\zeta) =        
\sum_{i=1}^{12} f_i(\zeta)$, where the functions $f_i(\zeta)$ are        
summarized in, e.g., eq. 20 of \citet{PW05}.         
         
The limits $\zeta_\pm(\alpha;\gamma,\alpha_1)$ depend on the energy           
of the outgoing photon, $\alpha$ for given $\alpha_1$ and $\gamma$          
\citep[see][]{PW05}.            
For $\alpha_1 \gg 1$ and $\gamma \simeq 1$, this dependence is          
degenerated and  $\zeta_\pm = 1 \pm \beta$.        
In this approximation, in calculating the difference of the 12        
function $\Delta f_i \equiv f_i(\zeta_+) - f_i(\zeta_-)$ only 4 out of        
the 12 terms are to the         
order $\alpha_1^{-1}$ or $\alpha^{-1}$, while           
the other 8 are to lower orders in $\alpha$, $\alpha_1$.          
Thus, to leading order in $\alpha, \alpha_1$,          
\beq          
\begin{array}{l}          
\Delta f_3 \equiv f_3(\zeta_+) - f_3(\zeta_-) \simeq {2 \beta \over          
(1-\beta^2) \alpha_1}, \\          
\Delta f_9 \simeq {2 \beta (1 + \beta^2) \over \alpha}, \\          
\Delta f_{10} \simeq {\beta^3 \over \alpha}, \\          
\Delta f_{11} \simeq {2 \beta (1 + \beta^2) \over \alpha^2}           
\left[ \alpha_1 - \alpha + \left( 1 + {\alpha_1 \over \alpha}          
\right)\left(1+ {\beta^2 \over 2}\right)\right].           
\end{array}          
\eeq          
            
The (normalized) energy loss rate of photon being scattered by unit density,          
isotropically distributed, mono-energetic electrons at Lorentz factor          
$\gamma$ is therefore given by             
\beq          
\frac{d\alpha(\gamma,\alpha_1)}{dt} =           
\int_{\alpha_{min}}^{\alpha_1}           
\alpha' \frac{d^2N(\gamma,\alpha_1)}{dt d\alpha'} d\alpha'          
\simeq          
c \sigma_T \left({1- \beta^2 \over 2} + {3 \beta^2 \over 16}\right) =           
c \sigma_T \left({1 \over 2} - {5 \beta^2 \over 16}\right) + O(\alpha_1^{-1}).          
\label{eq:dEdt}          
\eeq          
          
\begin{figure}          
\plotone{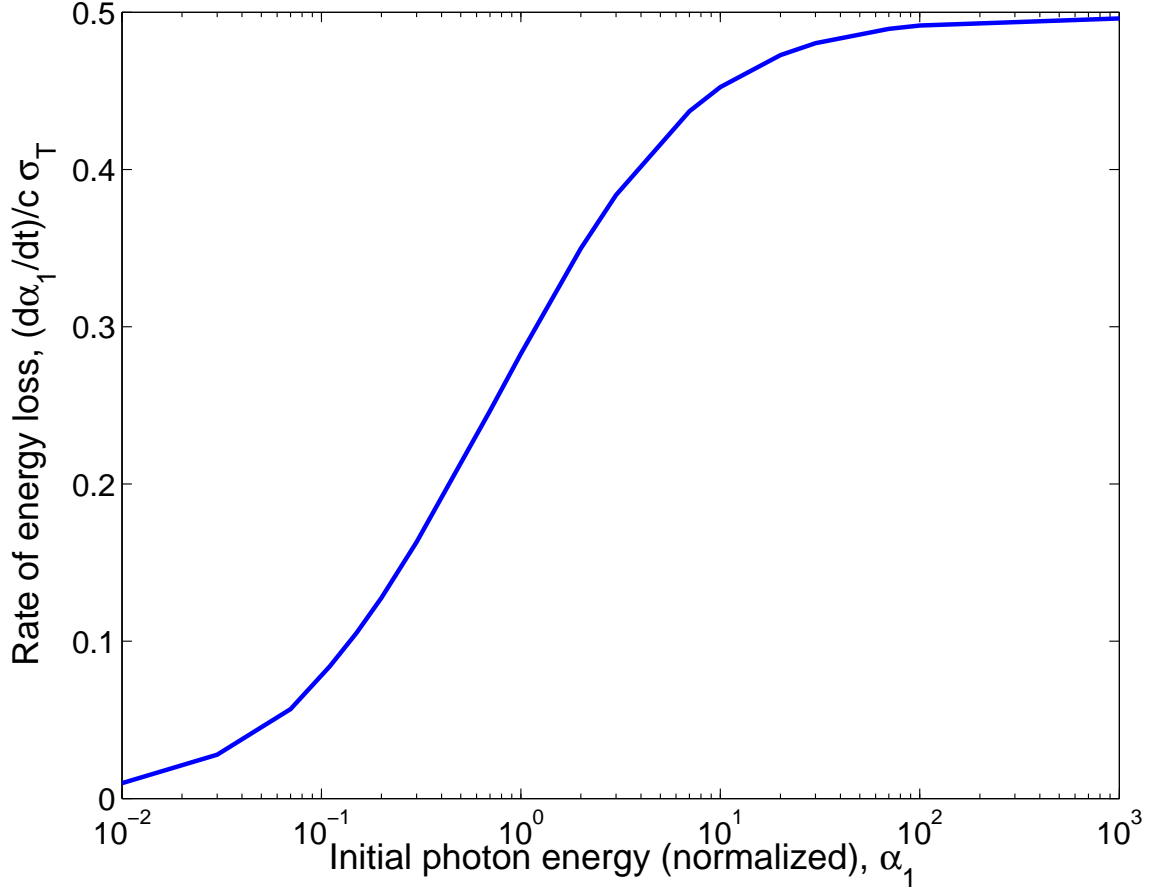}          
\caption{Energy loss rate of photon with initial energy $\alpha_1 m_e          
c^2$ due to Compton scattering with unit density, isotropically          
distributed, mono-energetic electrons with velocity $\beta = 0.1$.            
}          
\label{fig:e_loss_rate}          
\end{figure}

In order to check the validity of this approximation for values of          
$\alpha_1$ not much larger than 1, a numerical integration of          
equation \ref{eq:compton_rate_general} was carried out, using the          
exact formula. The results of the energy loss rate are presented in          
Figure~\ref{fig:e_loss_rate}.          
From the figure we find that the approximation $d\varepsilon/dt          
\approx c \sigma_T m_e c^2/2$, where $\varepsilon = \alpha m_e c^2$          
for the energy loss rate of photon due to Compton scattering with unit          
density, isotropically distributed, mono-energetic electrons with          
$\gamma \approx 1$ is valid for $\varepsilon \geq f m_e c^2$, with $f          
\approx 3$.             
          
\end{document}